\documentclass[AMA,LATO1COL]{WileyNJD-v2}
\articletype{Article Type}

\received{}
\revised{}
\accepted{}

\usepackage{type1cm}
\usepackage{siunitx}
\usepackage{float}
\usepackage{placeins}
\usepackage{xspace}
\usepackage{cleveref}
\usepackage{multirow}
\usepackage{subcaption}
\usepackage{graphicx}
\usepackage[mathlines]{lineno}

\graphicspath{{figures/}}

\def\etal{\textsl{et al.}}
\def\invivo{\textsl{in-vivo}}

\def\chem#1{\ensuremath{\rm #1}}
\def\T#1{\ensuremath{T_{#1}}\xspace}

\def\TR{\ensuremath{T_R}\xspace}

\def\TE{\ensuremath{T_E}\xspace}

\begin{document}

\title{Benchmarking GABA Quantification: A Ground Truth Data Set and Comparative Analysis of TARQUIN, LCModel, jMRUI and Gannet}

\author[1,2,3]{Christopher W Jenkins}
\author[4]{Max Chandler}
\author[4]{Frank C Langbein}
\author[1,2]{Sophie M Shermer}

\address[1]{College of Science (Physics), Swansea University, Swansea, SA2 8PP, United Kingdom}
\address[2]{Centre for Nanohealth \& Clinical Imaging Unit, Institute for Life Science, Swansea University, Swansea, SA2 8PP, United Kingdom}
\address[3]{Cardiff University Brain Research Imaging Centre (CUBRIC), Maindy Rd, Cardiff, CF24 4HQ, United Kingdom}
\address[4]{School of Computer Science and Informatics, Cardiff University, Cardiff, CF24 3AA, United Kingdom}

\corres{*S.\,M.~Shermer Swansea University \email{lw1660@gmail.com}}

\authormark{Jenkins \textsc{et al}}

\abstract[Summary]{
  Many tools exist for the quantification of GABA-edited magnetic resonance spectroscopy (MRS) data. Despite a recent consensus effort by the MRS community, literature comparing them is sparse but indicates a methodological bias. While \invivo{} data sets can ascertain the level of agreement between tools, ground-truth is required to establish accuracy, and investigate the sources of discrepancy.
  We present a novel approach to benchmarking GABA quantification tools, using several series of phantom experiments with iterated GABA concentration. Each series presents a different set of background metabolites and environmental conditions allowing comparison of not only individual estimates, but the ability of tools to characterise changes in GABA across a range of potential confounds. The methodology of the phantom experiments is presented, as well as characterisation of the data. We also perform an initial comparative analysis of several common MRS quantification tools (LCModel, TARQUIN, jMRUI, Gannet) and in-house code (LWFIT), to illustrate utility of the dataset and the potential bias introduced by different quantification methods. The GABA-to-NAA ratios reported by each tool are compared to the ground-truth, and estimation accuracy is assessed by linear regression of this relationship.
  While the linearity of GABA-to-NAA gradients is generally captured by all of the tools, a large variation in the slope, offset and environmental stability of the gradient is observed. The primary driver of differences in linear combination modelling is the choice of basis function. However, tools employing a common basis and pre-processing still produce differences on the order of $4\%$. Less-strictly parametrised fitting approaches appear to improve the robustness of quantification, but accurate modelling of edit efficiency calculations is still necessary to avoid systematic offsets. In general, the level of variation suggests that comparisons across quantification methods should be performed tentatively, and the sharing of basis sets and optimisation options is an important step towards greater reproducibility of \invivo{} MRS studies.
}

\keywords{MRS, GABA, Quantification, MEGA-PRESS, LCModel, TARQUIN, jMRUI, Gannet}

\maketitle

\section{Introduction}

Gamma-aminobutyric acid (GABA) is the primary inhibitory neurotransmitter in the central nervous system, maintaining the brain's excitation–inhibition balance. Its prominent role in neurotransmission and metabolism has led to extensive study and a plethora of applications for its detection. Disruption of GABAergic processes has been observed in Schizophrenic patients~\cite{Rowland2016, Simpson1989, Taylor2015} and GABA receptor dysfunction has associations with epilepsy~\cite{Leal2016, Pan2015}. GABA processes are prominent in type I diabetes~\cite{Soltani} and autism spectrum disorders~\cite{Brix2015}. GABA is also the subject of intense study by the psychology community, with GABA levels influencing impulsivity~\cite{Boy2011,Ende2016}, drug addiction~\cite{Cousins2002}, anxiety disorders~\cite{Robertson2016} and depression~\cite{Sanacora1999a}. The most common method for non-invasive detection of GABA \invivo{} is magnetic resonance spectroscopy (MRS). However, at clinically relevant field strengths the GABA signal is overlapped by more prevalent metabolites, necessitating other detection methods for reliable quantification.

While specialised techniques such as 2D $J$-spectroscopy~\cite{Ke2000} are used, the most common method is edited MRS~\cite{Harris2017, Mikkelsen2017, Mullins2014, Myers2016}, in particular MEGA-PRESS~\cite{Mescher1998, Mullins2014, Puts2012, Waddell2007}. MEGA-PRESS GABA-editing utilises two interleaved acquisitions --- `edit on' and `edit off' --- which differ only in the frequency of applied editing pulses. The edit frequencies are chosen to take advantage of the $J$-evolution of the target metabolites, such that upon subtraction of the two acquisitions, only resonances within the edit bandwidth and those coupled to them remain. The widespread use of edited MRS has seen the development of a range of software packages aimed at processing and quantification of the spectra~\cite{Edden2014, Naressi2001a, Provencher1993, Provencher2001, Reynolds2006, Stefan2009, Wilson2011a}. The tools process data by some combination of apodization, frequency and phase calibration, macromolecular baseline subtraction and residual water signal removal. Each step may be realised by a diverse array of methods, providing many potential sources of variation for the quantified signal. This effect is compounded by further differences in the quantification procedure. The most common approach to quantification is the generation of basis spectra --- via phantom scans or more commonly by quantum mechanical simulation --- followed by the application of a fitting algorithm to decompose the MRS signal into weighted combinations of the basis functions. For edited MRS, with its simplified spectra, other fitting approaches --- using Gaussian/Lorentzian modelling of resonances and numerical integration --- become more feasible.

A recent large-scale effort has been made to establish consensus in the MRS community~\cite{Consensus_Andronesi2020,Consensus_Choi2020,Consensus_Lin2021,Consensus_Near2020,Consensus_Oz2020,Consensus_Tkac2020,Consensus_Wilson2019}. While recommendations were made around the quantification of MRS data --- Choi~\etal{}~\cite{Consensus_Choi2020} reference basis set methods and Gaussian peak fitting for GABA quantification --- there was no consensus on the best approach. While it has been previously argued that different methods, suitably applied, exhibit similar variation and clinical observations~\cite{Mullins2003, Scott2016}, it is acknowledged in the literature that there is a procedural dependency in MRS analysis~\cite{zollner2021comparison,  Kanowski2004, Mandal2012, Poullet2008a,Mikkelsen2019}, which provides a statistically significant variation in reported metabolite amplitudes or relative concentrations~\cite{Bhogal2017, Marzola2014}. With the exact pre-processing, simulation and fitting parameters only partially reported in many research articles --- an issue discussed by Lin~\etal{}~\cite{Consensus_Lin2021} --- the methodological dependence of MRS quantification propagates, reducing the reproducibility of MRS studies. Despite this, there has been relatively little comparative study and benchmarking of the MRS analysis tools.

The goal of this study was to generate benchmarking data sets for the purpose of assessing the relative performance of GABA quantification tools and investigate the methodological dependence of GABA-edited data analysis. Previous work in this area has utilised either \invivo{} data --- for which there is usually no ground truth to evaluate the results --- or simulated data sets, which inevitably cannot capture all sources of experimental variation. For this study, MRS benchmarking data sets were generated using several series of calibrated phantom experiments. While phantoms fail to capture all influences of the \invivo{} environment, they do allow the inclusion of many experimental factors, while still enabling a ground truth comparison and ability to control the environmental characteristics expected to influence results. Four series of phantom data were experimentally acquired. In each case, the concentrations of various metabolites --- creatine (Cr), N-acetyl-L-aspartic acid (NAA), glutamine (Gln) and glutamate (Glu) --- are fixed at approximately \invivo{} levels, while the concentration of GABA is iteratively increased. At each GABA concentration step, MEGA-PRESS spectra were acquired. The four series of experiments were designed to increase the complexity of the spectra, starting with pH and temperature calibrated solutions containing only NAA, Cr and GABA (E1), then adding Glu and Gln to the solutions to increase the complexity of the spectra (E3), and finally progressing to a series of gel phantoms which more closely mimic \invivo{} spectra~\cite{portakal2018design} (E4\footnote{E4 corresponds to E4a in full data set released.}). To investigate the influence of the molecular environment, a series of non-neutral pH solutions of NAA, Cr and GABA (E2) were also acquired. The phantom data set has been designed with the benchmarking of new quantification tools in mind, for example in the use of convolutional neural networks for GABA quantification~\cite{mrsnet2019}. As such, the full data have been made available for others to benchmark their own analysis software, or optimise existing tools (DOI:10.21227/ak1d-3s20)~\cite{data}.

Furthermore, to demonstrate the utility of the dataset, MEGAPRESS spectra were quantified using a variety of state-of-the-art MRS analysis software to investigate the potential methodological bias that could be introduced. The GABA-to-NAA concentration ratio was estimated for each tool by applying appropriate signal attenuation corrections where needed. The reported ratios are plotted against the known ground-truth, and fit using linear regression. The GABA-to-NAA gradients are used to investigate the systematic offsets that can be introduced into reported concentration changes, and to inform future comparative benchmarking studies.

\section{Methods}

Phantom preparation procedures, experimental setup and scan parameters, and details of the analysis are described.

\subsection{Phantom preparation}

The phantom study consists of four experimental series --- E1, E2, E3 and E4 --- where the GABA (CAS-20791) concentration of a metabolite phantom is varied over multiple acquisitions using a fixed MRS protocol. The background metabolites --- NAA (CAS-997-55-7), Cr (CAS-6020-87-7), Glu (CAS-142-47-2) and Gln (CAS-56-85-9) --- are maintained at a fixed concentration for a given series, so that GABA gradients are purely a result of GABA concentration changes. E1, E2 and E3 are solution series, where the GABA concentration of a liquid phantom was increased incrementally. For series E4, several spherical gel phantoms were made with varying GABA concentration. The full concentration information of the experimental series is listed in \Cref{Tab:Conc} and representative images of the phantoms are shown in \Cref{Fig:PhantomPics}.

The solution series were prepared by dissolving the required concentration of background metabolites in de-ionised water to create a base solution. \SI{290}{\milli\litre} of the GABA-free base solution was transferred to a round bottom glass flask, filling it to the neck. To avoid dilution of the background metabolites, a small amount of this base solution was removed from the flask using a syringe and GABA was added to prepare a concentrated GABA solution (\SI{1}{\milli\gram/\milli\litre}). To vary the GABA concentration between scans, a small amount of the solution was removed from the flask with a syringe and replaced by the same volume of the concentrated GABA solution. The amount of solution replaced in each step varied depending on the desired GABA concentration change, but was generally between \SI{0.5}{\milli\litre} and \SI{2}{\milli\litre}. This procedure was repeated several times, iterating GABA to a relatively high concentration with the same scan protocols applied for each concentration step. This procedure limited experimental errors solely to the GABA concentration. The pH of the solution was monitored between scans and adjustments were made using a $36\%$ hydrochloric acid (CAS-7647-01-0) solution and a $3.99\%$ sodium hydroxide solution (CAS-1310-73-2) to maintain a pH of $7.2 \pm 0.2$. The amounts of HCl or NaOH added after the initial pH calibration of the base solution were marginal ($\le \SI{1}{\milli\litre}$ for the \SI{290}{\milli\litre} flask) and any dilution of the overall solution was deemed negligible. The E2 series was one exception to this procedure, where the pH was intentionally not adjusted, to investigate the effect this had on the various quantification procedures. The final pH of E2 was measured to be $3.0 \pm 0.2$.

For the gel series, the phantoms were made in advance of scanning. \SI{800}{\milli\litre} of a base solution of Glu, Gln, NAA and Cr was prepared and divided into eight \SI{100}{\milli\litre} portions. Different amounts of GABA were added to each solution and pH calibration was performed as above. Finally, \SI{1}{\gram} of agar (CAS-9002-18-0)~\footnote{Food-grade agar was sourced for this study} was added as gelling agent. The mixtures were then heated to \SIrange{90}{100}{\degreeCelsius} while being stirred until the agar had fully dissolved. A small hole ($<\SI{2}{\milli\metre}$ diameter) was created in a spherical plastic mould using a plastic welding tool and the solutions were injected via a syringe. The arrangement was then allowed to cool over night. Once solidified, the gels were examined and the opening was sealed using a small amount of silicone sealant. Spherically-shaped phantoms were deemed the most suitable to reduce magnetic susceptibility-induced field inhomogeneity and minimise the spectral linewidth.

The accuracy of the scale used for mass measurements is \SI{1}{\milli\gram} and the graduated cylinders used for the volume measurements have an accuracy of \SI{1}{\milli\litre}. Assuming negligible errors in the manufacturer's stated molar masses, the uncertainty $\Delta c$ in the molar concentration $c = \frac{m}{MV}$, where $m$ is the mass in grams, $M$ is the molar mass of the solute and $V$ is the volume of the solvent, is $\frac{\Delta c}{c} = \sqrt{\left[\tfrac{\Delta m}{m}\right]^2 + \left[\tfrac{\Delta V}{V}\right]^2}$. Assuming (generously) an uncertainty of $2\%$ for the mass measurements and $1\%$ for the volume measurement, the uncertainty in the concentrations of base metabolites is estimated to be less than $3\%$. The uncertainty in the amount of solution added or removed from the flask in each step via \SI{5}{\milli\litre} syringes is slightly higher and could be up to $10\%$, assuming an uncertainty of \SI{0.1}{\milli\litre} per \SI{1}{\milli\litre} to account for slight misreadings of the plunger position and tiny air pockets within the syringe.

\begin{table}[!t]
  \centering
  \caption{Concentration information in units of $\si{mM} = \si{\milli\mole/\litre}$. All series, except E2, are pH calibrated to $7.0 \pm 0.2$.}\label{Tab:Conc}
  \scalebox{1}{\begin{tabular}{|c|cccc|l|} \hline
    Series    & NAA   & Cr     & Glu    & Gln   & GABA\\\hline
    E1  & $15.00$ & $8.00$ &  $0.00$ & $0.00$ & $0.00$, $0.52$, $1.04$, $1.56$, $2.07$,  $2.59$, $3.10$, $4.12$, $6.15$, $8.15$, $10.12$, $11.68$\\
    E2 & $15.00$ & $8.00$ & $0.00$ & $0.00$ & $0.00$, $0.50$, $1.00$, $1.50$, $2.00$, $2.50$, $3.00$, $3.99$, $4.98$, $5.96$, $6.95$, $7.93$, $8.90$, $9.88$, $11.81$\\
    E3 & $15.00$ & $8.00$ & $12.00$ & $3.00$ & $0.00$, $1.00$, $2.00$, $3.00$, $3.99$,  $4.98$, $5.97$, $6.95$, $7.93$, $8.91$, $9.88$, $10.85$, $11.81$, $12.77$, $13.73$\\
    E4  & $15.00$ & $8.00$ & $12.00$ & $3.00$ & $0.00$, $1.00$, $2.00$, $3.00$, $4.00$, $6.00$, $8.00$, $10.00$\\\hline
  \end{tabular}}
\end{table}

\subsection{Scan Protocols and Data Acquisition}

All MR scans were conducted at Swansea University's Clinical Imaging Facility using a MAGNETOM Skyra \SI{3}{T} (Siemens Healthcare GmBH, Erlangen, Germany). The scanner room was temperature controlled to $\SI{20}{\pm 0.6\degreeCelsius}$. Between each scan, the phantom was removed to update concentrations, so the influence of heating is deemed negligible. Signal acquisition was done using the built-in spine coils, specifically the four channel spine coil element `SP2'. The spine coils provided the platform for the most reproducible set-up and SP2 exhibited the highest SNR of the coil elements available. The phantom was aligned with this element and raised to isocenter, using a cardboard phantom holder to ensure maximum field homogeneity for the scans. Double-spin-echo field maps were acquired for each phantom to assess the homogeneity, and manual shimming and frequency calibration was performed to optimise the spectral width. Further manual calibration was performed to adjust the TX reference voltage and water-suppression pulses.

A $20\times20\times20 \si{\cubic\milli\metre}$ voxel at the isocenter of the magnet was selected for single voxel spectroscopy. The GABA-edited spectra were acquired using the MEGA-PRESS pulse sequence WIP859G~\footnote{Work in Progress. The product is currently under development and it is not for sale in the US and in other countries. Its future availability cannot be assured.} with CHESS water-sup\-pres\-sion~\cite{Ogg1994}, $\TR = \SI{2000}{\milli\second}$, $\TE = \SI{68}{\milli\second}$, $N=160$ averages and a sampling frequency of $\SI{1250}{\hertz}$, $N=2048$ samples. Editing pulses were applied at $\SI{1.9}{ppm}$ during the on acquisition and $\SI{7.4}{ppm}$ during the off acquisition. Spectra with water suppression turned off were also acquired to calculate linewidth, water suppression factors and to facilitate eddy current correction and water concentration referencing.

For each experiment, the raw data acquired, the single-average coil-combined spectra and combined average edit-on, edit-off and difference spectra produced by the vendor-supplied spectroscopy software have been made available~\cite{data}. For the purpose of this paper, the combined channel and average dicom spectra were used. While raw data is generally preferable in order to appraise frequency and phase drift correction, the averaged dicom data was most universally supported by all the software packages evaluated, allowing a more encompassing cross-section of the methods. All spectra used in this study were inspected for reconstruction issues, prior to their inclusion.

\subsection{Spectral quality}

Prior to quantification, the \SI{2}{ppm} NAA and water peaks in each spectrum are fitted by Lorentzians, from which the position, full-width-half-maximum (FWHM) and maximum signal amplitude of the resonances are obtained. The noise floor was estimated by computation of the standard deviation, $\sigma$, of the signal in the region $8-$\SI{9}{ppm}, observed to be free of metabolite signal, then the signal-to-noise ratio (SNR) calculated: $\frac{S}{2\sigma}$. Water suppression factors are estimated by comparing the height and area of the water peak in the water unsuppressed and water-suppressed spectra.

\subsection{Analysis methods}

Analysis methods can be broadly separated into two categories: (i) basis set methods (TARQUIN~\cite{Wilson2011a}, jMRUI~\cite{Naressi2001a,Stefan2009}, LCModel\cite{Provencher2001}), which attempt to fit the data using a set of simulated or experimentally obtained basis spectra and (ii) peak fitting and integration tools (Gannet\cite{Edden2014}, LWFIT -- An in-house peak integration method used as a baseline in this study), which compute the area of peaks associated with various metabolites. As the absolute amplitudes are arbitrary and depend on the scanner design, transmitter calibration, coils used and the phantom itself. We focus on quantifying the relative contributions of metabolite signals, specifically the GABA-to-NAA ratio. The concentration ratio estimate, $C$, between a target metabolite, $M$, and reference, $R$, is given by:
\begin{equation}
  C = \frac{S_{M}}{S_{R}} \frac{N_R R_R E_R}{N_M R_M E_M},
\end{equation}
where $S$ is the reported resonance amplitude, $N$ is the number of contributing spins, $R$ is the signal attenuation factor resulting from $\T1$ and $\T2$ relaxation, and $E$ is the edit efficiency. For the purposes of our analysis, the difference between the relaxation properties of GABA and NAA are assumed negligible, as suggested by Choi~\etal~\cite{Consensus_Choi2020}. With this assumption, basis set methods report appropriately scaled metabolite amplitudes. However, the peak fitting methods require corrections for $N$ and $E$. Both Gannet and LWFIT quantify the \SI{3}{ppm} GABA resonance and \SI{2}{ppm} NAA singlet, resulting in $\frac{N_R}{N_M} = 3/2$. By default, Gannet's edit efficiency correction factor is estimated as $E_R/E_M = 0.5$, and LWFIT's correction factor was calculated from simulation (\Cref{Results_Simulations}). The concentration ratios reported by each tool were plotted vs. the known ground-truth ratio, and the relationship calculated via linear regression.

For the purposes of this study, the tools were employed as close to ``out-of-the-box'' as possible. Default settings were used with additional developer recommendations whenever these were readily available. While it is indeed possible to further optimise the parameters of each tool, our aim was to examine the potential methodological variation that could occur in GABA MRS studies. A brief description of each tool is provided below, and readers are directed to the referenced literature for further information.

\subsubsection{Gannet}

GABA-MRS analysis tool (Gannet)~\cite{Edden2014} is an open-source automated MATLAB-based analysis tool, specifically designed for automated processing of \SI{3}{\tesla} GABA MEGA-PRESS data. Gannet is able to process raw data, further reducing user dependence, performing channel combination, adding line broadening, frequency and phase corrections, outlier rejection and time averaging. Fitting is performed using non-linear least-squares algorithms (lsqcurvefit and nlinfi) with a five parameter Gaussian model to estimate the $\SI{3}{ppm}$ GABA signal in the difference spectrum, a six parameter Lorentzian model to estimate the $\SI{3}{ppm}$ Cr signal in the off spectrum and, if appropriate, a six parameter Gaussian-Lorentzian model to fit the unsuppressed water signal. Quantitative results are reported as integral ratios of GABA to Cr and a concentration relative to water, NAA or Glx (Glu and Gln combined).  While Gannet attempts to remove user interaction from the analysis procedure, it allows user modification of the code and adjustment of the pre-initialisation script is encouraged depending on setup. For this paper Gannet version 3.0 was used and the pre-initialisation script was adjusted to account for room temperature phantom acquisition.

\subsubsection{jMRUI}

jMRUI~\cite{Naressi2001a, Stefan2009} is proprietary software distributed under its own license terms and made freely available to registered users for non-commercial use. It has a large suite of pre-processing, analysis and simulation options. For this study, two basis set methods were used to quantify the spectra. The first, quantitation based on semi-parametric quantum estimation (QUEST)~\cite{Ratiney2004,Ratiney2005}, is a time-domain fitting tool, which includes a semi-parametric approach to handle spurious signals resulting from macro molecules and lipids. The second method, automated quantitation of short echo time MRS spectra (AQSES)~\cite{Poullet2007a}, uses a modified VARPRO variable projection NLLS algorithm~\cite{Sima2007}, which imposes prior knowledge in the form of upper and lower bounds on the non-linear parameters.

For this study, jMRUI version 6.0 beta was used. MEGA-PRESS basis sets were generated using NMR-SCOPE-B~\cite{Starcuk2009, Starcuk2017} by defining a single MEGA-PRESS pulse sequence with sequence parameters matching our experimental protocol. Two basis sets for GABA/Cr/NAA and GABA/Cr/NAA/Glu/Gln were generated. Manual frequency and phase calibration and apodization were performed on the input spectra and fitting with both tools was applied using this common input. This allowed investigation of the influence of the fit algorithm itself, independent of pre-processing.

\subsubsection{LCModel}

LCModel~\cite{Provencher1993, Provencher2001} is a formerly commercial, Linux-based MRS analysis tool which recently moved to open-source. It can be used with its internal \emph{in-vitro} basis set or any arbitrary basis set specified by the user. The baseline signal resulting from macromolecules is established using spline fits. Fitting is performed using a Gauss-Newton non-linear least-squares algorithm with Marquardt modification~\cite{Marquardt1963} with additional terms for $T_2$, spectral shift and field inhomogeneities. For this paper, version 6.3-1L and the Purcell lab basis set~\cite{Dr.JimMurdoch} recommended by LCModel's creators were used.

\subsubsection{LWFIT}

Our in-house code~\cite{Shermer2021}, LWFIT, written in MATLAB, aligns the edit-off and difference spectra so that the main NAA peak is precisely at \SI{2.01}{ppm} and then calculates the areas for the NAA (\SIrange{1.91}{2.11}{ppm}, difference), Cr (\SIrange{2.90}{3.10}{ppm}, edit-off) and GABA (\SIrange{2.90}{3.12}{ppm}, difference) peaks, as well the two main peaks associated with Glu and Gln, GLX1 (\SIrange{2.25}{2.45}{ppm}, difference) and GLX2 (\SIrange{3.65}{3.85}{ppm}, difference), using the real frequency-domain spectra. This is done by numerical integration using piecewise-linear functions over the indicated fixed \si{ppm} ranges, selected to minimise contamination from other signals. Lorentzian, Gaussian and spline fitting and several baseline fitting and filtering methods were explored but initial testing indicated that numerical integration with minimal pre-processing yielded the most accurate estimation results of the methods considered. While more aggressive noise filters aesthetically improve the quality of the spectra, their application tends to exacerbate underestimation of GABA, especially for weak signals, suggesting that filtering eliminates some of the GABA signal present in the data~\cite{PhD_Jenkins}. For this work, only zero-filling ($N=4096$ points) and a frequency shift to align the NAA peak in the spectrum to \SI{2.01}{ppm} are performed.

\subsubsection{TARQUIN}

Totally Automatic Robust Quantitation in NMR (TARQUIN)~\cite{Reynolds2006, Wilson2011a} is a cross platform, time domain basis set analysis tool. TARQUIN is an open source software package written in C++, complete with a GUI (Graphical User Interface) and a built-in NMR simulator. TARQUIN uses a Lawson-Hanson non-negative least-squares algorithm~\cite{Lawson1995} with a basis set of pre-simulated spectra, which can be provided by the user. Residual water removal is performed using HSVD (Hankel Singular Value Decomposition)~\cite{Barkhuijsen1987, deBeer1992} and automated phase and frequency correction is applied. TARQUIN does not perform a full MEGA-PRESS simulation, but rather models the expected signal as a pseudo-doublet of Gaussian peaks and adjusts its phase correction procedures to accommodate the negative NAA peak. Frequency calibration is made relative to NAA. For this paper, TARQUIN version 4.3.11 with its internal MEGA-PRESS basis set was used, and fit parameters were adjusted for edited data, as directed by the user manual.

\subsection{Simulations}

To further facilitate interpretation of the experimental results, quantum mechanical simulations were also performed. Simulation is fundamental in MRS quantification. Basis set fitting methods utilise simulated spectra to model the peak structure to which the experimental data are fit. While peak fitting methods are less tightly parametrised during the fit, they still must compensate for edit efficiency -- the proportion of peak amplitude lost as a part of the editing process -- in order to accurately quantify the metabolites. The simulations were used to investigate the influence of ideal RF-pulse simulation and GABA models on quantification, and to facilitate edit efficiency correction for LWFIT.

The experimental setup was emulated, modulating GABA concentration while maintaining a fixed concentration of other selected metabolites. MEGA-PRESS simulations were performed using FID-A~\cite{Simpson2017}, a simulation and data processing package for MRS. MEGA-PRESS difference spectra were generated, mirroring the sequence parameters of the phantom study with a finite-bandwidth editing pulse alternating between $\SI{1.9}{ppm}$ and $\SI{7.4}{ppm}$ at a field strength of $\SI{2.89}{\tesla}$ with an acquisition bandwidth of $\SI{1250}{\hertz}$ and two and four step phase cycling. Spectra were generated for GABA, Cr, NAA, Glu and Gln. The difference spectra were combined according to the concentrations used in the phantom experiments. The metabolite models simulated were derived from the work of Govindaraju \etal{}~\cite{Govindaraju2000}. For GABA, there is a greater degree of uncertainty~\cite{Kreis2012} and basis sets were generated for three distinct models that have been popularised in the literature by Govindaraju \etal{}~\cite{Govindaraju2000}, Kaiser \etal{}~\cite{Kaiser2008} and Near \etal{}~\cite{Near2013}, respectively. Simulated series were tested with all three GABA models to investigate any potential bias of our simulated data. For all simulated series, GABA concentrations were varied between $\SI{1}{mM}$ and $\SI{12}{mM}$ at $\SI{1}{mM}$ intervals and the combined spectra were analysed using our in-house code.

\section{Results and Analysis}

We initially present the experimentally acquired spectra along with the corresponding calibration and quality measures. We then present the simulation and quantification results.

\subsection{Calibration and characterisation of spectra}

The stability of the experimental setup over the course of an entire series of experiments was assessed by considering the variation of core parameters such as transmitter voltage and frequency, shim gradients and water suppression settings. \Cref{tab:calibration} shows that a small adjustment made to the experimental procedure for E3 and E4 to avoid table movements between successive scans significantly reduced adjustments of transmitter frequency, reference voltage and linear gradient offsets $G_x$, $G_y$, $G_z$, although all results are within acceptable limits.

\begin{figure}
  \centering
  \begin{subfigure}{0.45\linewidth}
    \includegraphics[width=1\linewidth]{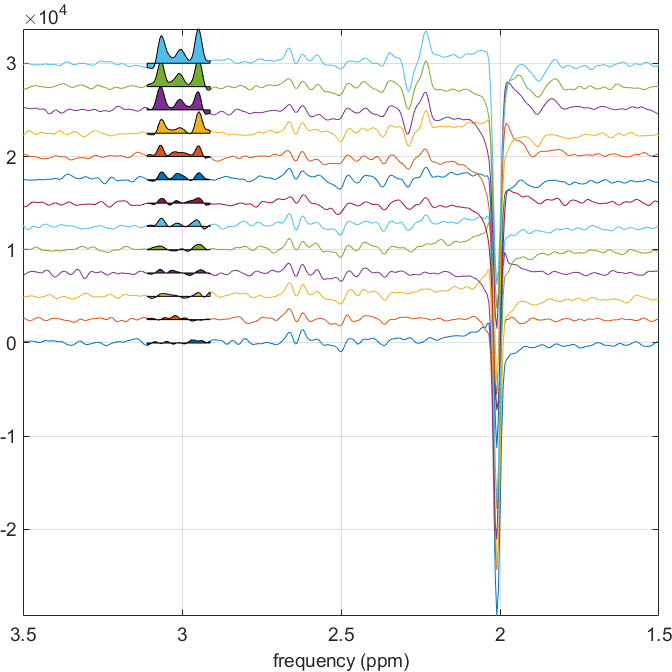}
    \subcaption{Spectra for series E1}
  \end{subfigure} \hfill
  \begin{subfigure}{0.45\linewidth}
    \includegraphics[width=1\linewidth]{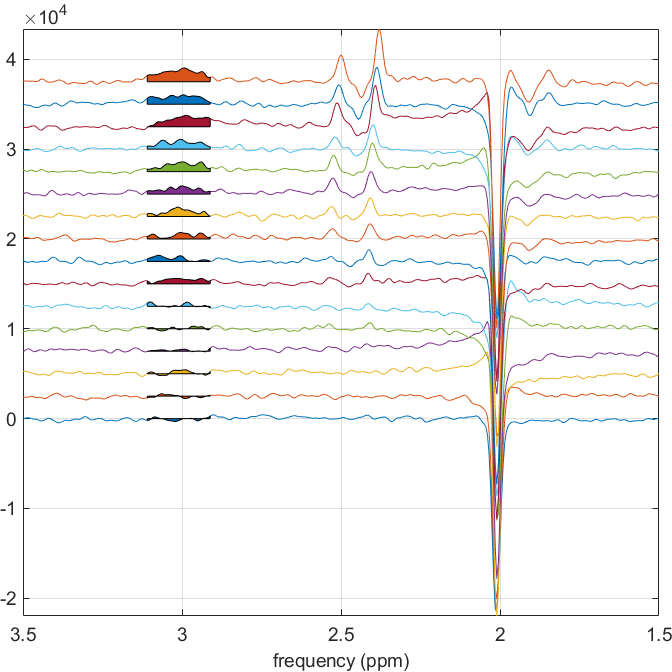}
    \subcaption{Spectra for series E2}
    \end{subfigure} \\
  \begin{subfigure}{0.45\linewidth}
    \includegraphics[width=1\linewidth]{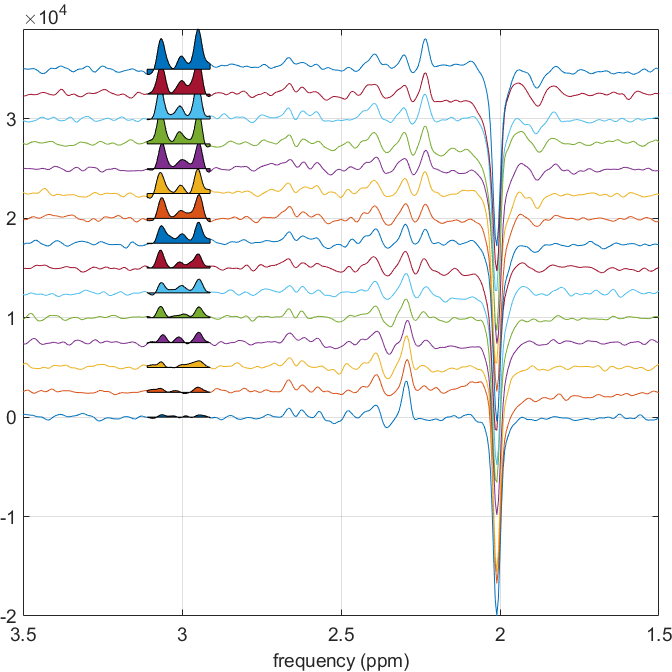}
    \subcaption{Spectra for series E3}
  \end{subfigure} \hfill
  \begin{subfigure}{0.45\linewidth}
    \includegraphics[width=1\linewidth]{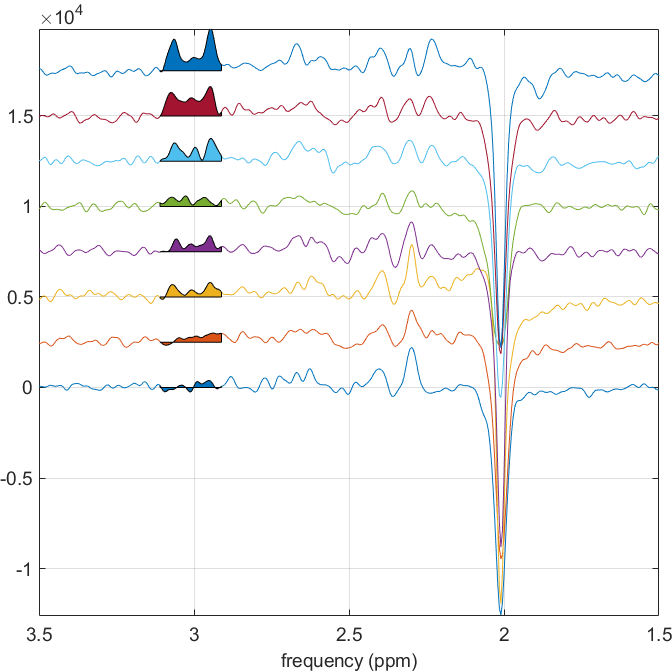}
    \subcaption{Spectra for gel phantom series E4}
  \end{subfigure} \\
  \caption{Difference spectra aligned to \SI{2.01}{ppm} NAA peak. The \SI{3}{ppm} GABA peak is shaded according to LWFIT integration limits, and vertically offset spectra correspond to increasing concentrations of GABA for the phantoms.}\label{fig:spectra}
\end{figure}

The difference spectra for all four experimental series are shown in \Cref{fig:spectra}. The expected pseudo-triplet structure, clearly visible for the well-calibrated solutions E1 and E3, is still discernible in the gel phantom spectra E4, despite the broader linewidth. However, this structure is not present in E2 where we observe a broad, almost Gaussian resonance. Despite this, the highlighted area does appear to increase linearly with GABA concentration.

\Cref{table:FWHM_WS_SNR} gives the linewidths of the water and NAA peaks obtained from Lorentzian peak fits, indicative values for the water suppression factors (WSF) and the SNR of the NAA peak. While the linewidths increase for the gel phantoms as expected, the width of the NAA peak in the difference spectrum remains around \SI{4}{\hertz}, suggesting good field homogeneity and frequency stability over the duration of the experiments. SNR over $200$ for NAA in the solution series and over $100$ for the gel phantoms are also well within the acceptable quality limits.

\begin{table}[!t]
  \caption{Total variation ($\Delta$) and standard deviation ($\sigma$) of transmitter frequency $\nu$ in \si{\hertz}, reference voltage TX in Volts, and linear gradient offsets $G_x$, $G_y$, $G_z$ for the four experimental series.} \label{tab:calibration}
  \centering
  \begin{tabular}{|l|cc|cc|cc|cc|cc|} \hline
    & $\Delta\nu$  & $\sigma(\nu)$ & $\Delta \mbox{TX}$ & $\sigma(\mbox{TX})$
    & $\Delta G_x$ & $\sigma(G_x)$ & $\Delta G_y$ & $\sigma(G_y)$ & $\Delta G_z$ & $\sigma(G_z)$\\\hline
    E1 & $22$ & $6.02$ & $13.4$ & $4.41$ & $37$ & $9.25$  & $92$ & $33.13$ & $19$ & $7.03$  \\
    E2 & $28$ & $7.42$ & $17.1$ & $4.89$ & $36$ & $10.58$ & $51$ & $16$    & $62$ & $16.52$ \\
    E3 & $3$  & $0.77$ & $0.8$  & $0.32$ & $10$ & $3.20$  & $37$ & $7.25$  & $85$ & $28.25$ \\
    E4 & $9$  & $3.18$ & $0.3$  & $0.11$ & $0$  & $0$     & $0$  & $0$     & $0$  & $0$     \\\hline
  \end{tabular}
\end{table}

\begin{table}
  \caption{Linewidths (mean/std over series in \si{\hertz}) of \chem{H_2O} peak in spectra acquired without water suppression (WS OFF) and the EDIT ON/OFF spectra with water suppression, NAA peak for EDIT ON and difference (DIFF) spectra with water suppression, as well as median water suppression factor (WSF) and SNR of NAA peak.}
  \centering
  \label{table:FWHM_WS_SNR}
  \scalebox{0.99}{
    \begin{tabular}{|l|ccc|c|cc|c|}\hline
      &\chem{H_2O} WS OFF& \chem{H_2O} EDIT ON & \chem{H_2O} EDIT OFF & WSF    & NAA EDIT OFF    & NAA DIFF        & SNR NAA\\\hline
      E1 & $2.40 \pm 0.67$  & $2.10 \pm 0.54$     & $2.11 \pm 0.55$      & $1840$ & $1.19 \pm 0.31$ & $1.19 \pm 0.31$ & $209.6$\\
      E2 & $3.04 \pm 1.44$  & $2.82 \pm 1.03$     & $2.82 \pm 1.03$      & $1873$ & $1.84 \pm 0.33$ & $1.85 \pm 0.33$ & $223.0$\\
      E3 & $2.98 \pm 0.90$  & $4.16 \pm 2.02$     & $4.16 \pm 2.01$      & $665$  & $1.97 \pm 0.25$ & $1.98 \pm 0.25$ & $222.0$\\
      E4 & $11.16\pm 3.19$  & $12.42\pm 9.01$     & $12.54 \pm 8.91$     & $478$  & $4.20 \pm 1.48$ & $4.25 \pm 1.55$ & $109.7$\\\hline
  \end{tabular}}
\end{table}

\subsection{Simulation results}\label{Results_Simulations}

\begin{figure}[!ht]
	\centering\includegraphics[width=0.65\linewidth]{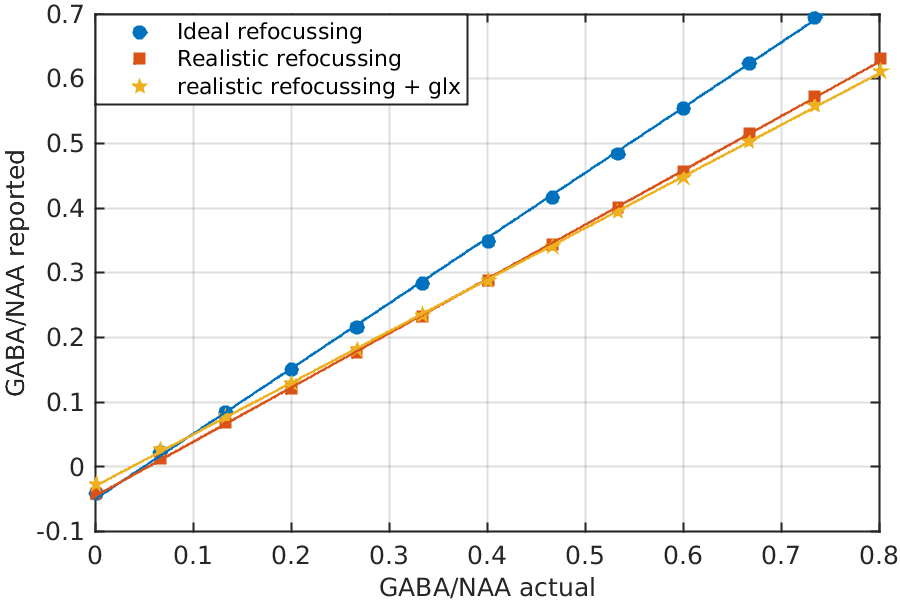} \\
	\centering\scalebox{0.9}{\begin{tabular}{|c|c|c|c|}
			\hline
			Series    & Gradient fit (95\% CI)   & Intercept (95\% CI) & $R^2$\\
			\hline
			Ideal      & $1.01$  ($1.00$, $1.02$) & $-0.050$  ($-0.055$, $-0.044$)  & $1.00$ \\
			Realistic  & $0.84$  ($0.83$, $0.85$) & $-0.045$  ($-0.048$, $-0.042$) & $1.00$ \\
			Realistic \& Glu+Gln & $0.80$  ($0.80$, $0.80$) & $-0.029$  ($-0.032$, $-0.027$) & $1.00$ \\
			\hline
	\end{tabular}}
	\caption{GABA quantification results (LWFIT) for series of simulated spectra and linear fit parameters with 95\% confidence intervals.}\label{fig:s1}
\end{figure}

\begin{figure}[!ht]
	\includegraphics[width=0.32\linewidth]{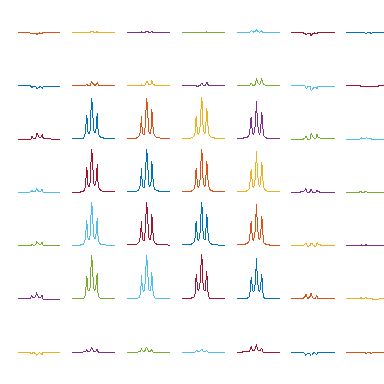}
	\includegraphics[width=0.32\linewidth]{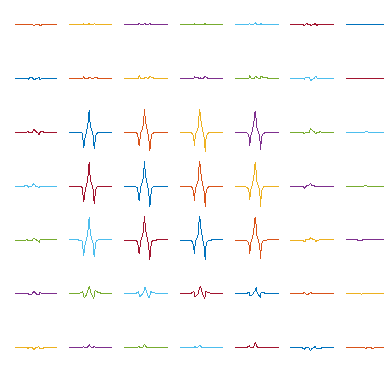}
	\includegraphics[width=0.32\linewidth]{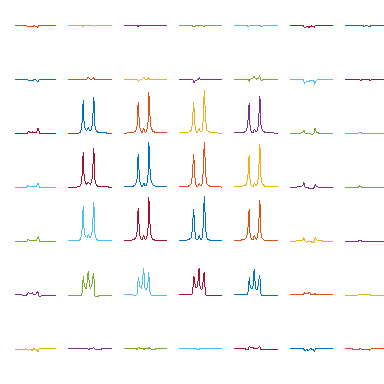}
	\caption{Edit on, edit off and difference spectra for GABA (Near model). 2D simulations performed using FID-A on an $8\times 8$ grid over a $\SI{5}{\centi\metre} \times \SI{5}{\centi\metre}$ region with a target excitation volume of $\SI{2}{\centi\metre}\times\SI{2}{\centi\metre}$ in the centre. Shaped refocusing pulses significantly reduce the GABA editing efficiency near the voxel boundary, resulting in underestimation of GABA if not taken into account.}\label{fig:Sim_2D}
\end{figure}

\Cref{fig:s1} shows the GABA-to-NAA amplitude ratio obtained by LWFIT vs. the actual imposed GABA-to-NAA ratio for three series of simulated experiments. Simulations using ideal excitation and refocusing pulses produce a gradient of $1.01$ --- within $1\%$ of the actual ratio --- an intercept of $-0.05$ and an $R^2$ value of $1.00$ (up to $3$ significant digits). However, for spectra simulated with finite-duration refocusing pulses with realistic pulse envelopes, the GABA-to-NAA ratio gradient is only $84\%$ of its expected value for the NAA-GABA-Cre series and this is further reduced to $80\%$ when Glu and Gln are added. \Cref{fig:Sim_2D} shows the \SI{3}{ppm} GABA peak for the edit-on, edit-off and difference spectra as a function of position on a 2D grid. The difference spectra near the boundary of the selected voxel indicate that the editing efficiency is significantly reduced. In an experimental setting, the editing efficiency will be further reduced by imperfect slice profiles of the excitation pulses and other factors such as field inhomogeneity. Assuming these issues lead to a further reduction of the editing efficiency by a factor of $0.84$ and $0.80$, respectively, the simulations predict GABA-to-NAA ratio gradients of $0.7$ and $0.64$ of the actual value for E1 and E3, respectively. These values are used as correction factors for LWFIT's quantification of the experimental series.

Edit efficiency is a crucial factor in edited MRS quantification. Basis set fitting methods should have this incorporated into the basis by appropriate simulation of the basis functions. Peak fitting methods, on the other hand, require an ad-hoc factor to compensate for the resulting reduction in peak amplitude. To ensure the best GABA estimates, full simulations with correct pulse envelopes are required to fully capture the $J$-modulated peak structure for basis set methods, and editing efficiency factor for peak fitting.

\subsection{Quantification}

\begin{figure}[!t]\hfill
  \begin{subfigure}{.49\textwidth}
    \includegraphics[width=1\linewidth]{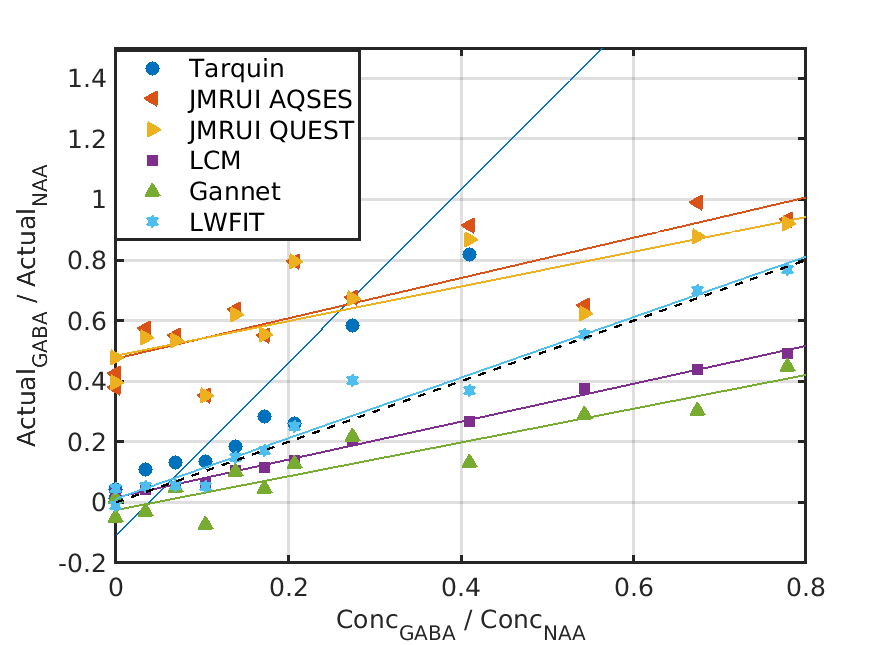}
    \subcaption{E1 phantom series results.}
    \label{fig:E1}
  \end{subfigure}\hfill
  \begin{subfigure}{.49\textwidth}
    \includegraphics[width=1\linewidth]{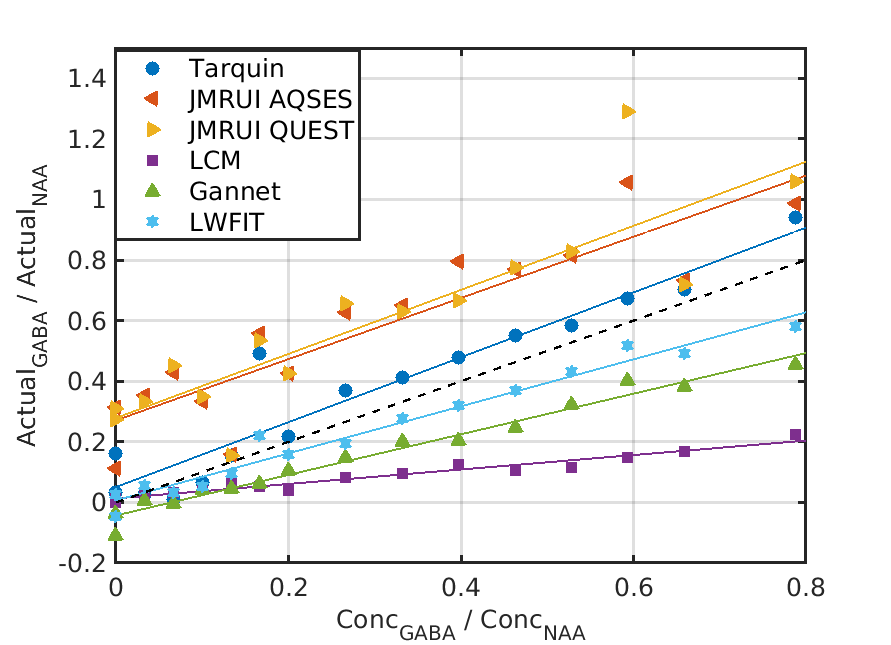}
    \subcaption{E2 phantom series results.}
    \label{fig:E2}
  \end{subfigure}\\
  \begin{subfigure}{.49\textwidth}
    \includegraphics[width=1\linewidth]{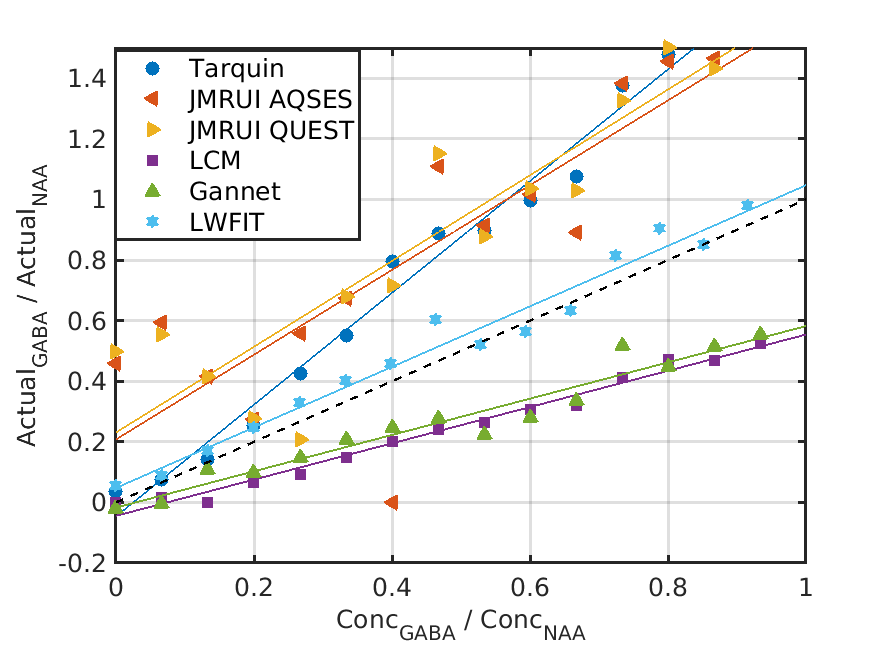}
    \subcaption{E3 phantom series results.}
    \label{fig:E3}
  \end{subfigure}\hfill
  \begin{subfigure}{.49\textwidth}
    \includegraphics[width=1\linewidth]{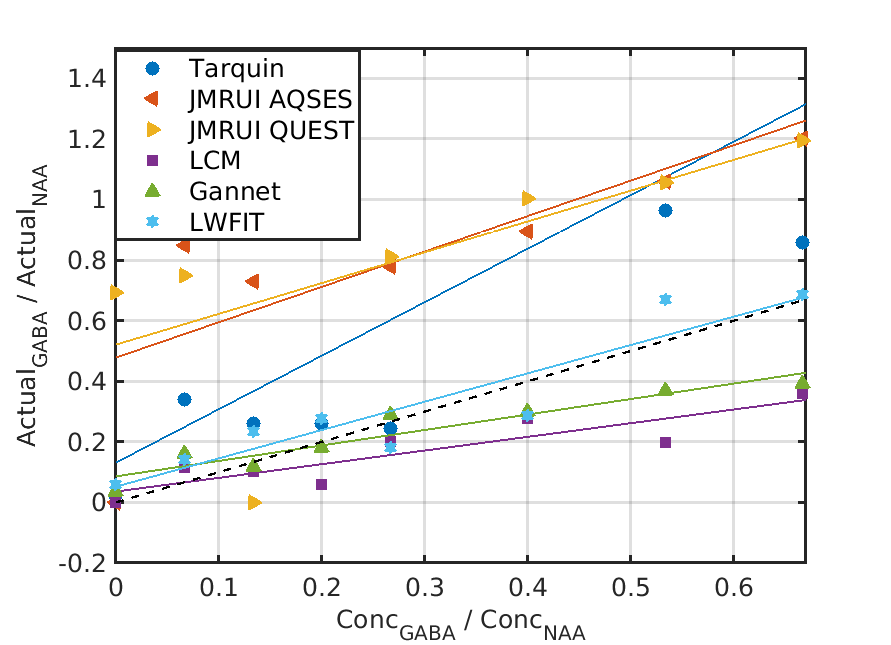}
    \subcaption{E4 phantom series results.}
    \label{fig:E4}
  \end{subfigure}\\
  \caption{GABA-to-NAA amplitude ratios reported by different tools vs. actual concentration ratio. In each plot, a black hashed line is used to represent the ideal case of $x=y$.}
\end{figure}

\begin{table}
  \caption{Linear fit results for GABA-to-NAA ratios in all four experimental series. Reported are the gradient $a$ and intercept $b$ of the linear fit with $95\%$ confidence intervals (CI) and the coefficient of determination $R^2$ (to two significant digits). Perfect quantification corresponds to $a=1$, $b=0$ and $R^2=1$. The percentage variation of reported amplitudes for NAA, whose actual concentration is constant, is also given. The large $\%$ variation of NAA amplitudes reported by TARQUIN for E1 is due to an unexplained rescaling of metabolite amplitudes between scans $10$ and $11$. The $\%$ variation of NAA for scans 1 -- 10 is $6.09\%$. There is no simple explanation for the large variation reported by TARQUIN for E4.}
  \label{Tab:Experimental}
  \centering
  \begin{tabular}{|c|l|ll|c|c|}
    \hline
    Series & Tool & Gradient $a$ with $95\%$ CI & Intercept $b$ with $95\%$ CI & $R^2$ & NAA$\%$\\
    \hline
    E1 & TARQUIN   & 2.85 (2.41, 3.30)    & -0.11 (-0.27, 0.053)       & 0.95 & 38.20\\
    & jMRUI AQSES  & 0.67 (0.37, 0.97)    & 0.48 (0.37, 0.58)          & 0.68 & 3.86\\
    & jMRUI QUEST  & 0.57 (0.30, 0.85)    & 0.48 (0.38, 0.58)          & 0.66 & 4.03\\
    & LCModel      & 0.63 (0.60, 0.66)    & 0.016 (0.0060, 0.027)      & 1.00 & 2.76\\
    & Gannet       & 0.56 (0.42, 0.70)    & -0.025 (-0.076, 0.026)     & 0.87 & 3.39\\
    & LWFIT        & 0.98 (0.87, 1.10)    & 0.012 (-0.029, 0.053)      & 0.97 & 2.61\\
    \hline
    E2 & TARQUIN   & 1.07 (0.87, 1.27)    & 0.051 (-0.026, 0.13)       & 0.90 & 2.87\\
    & jMRUI AQSES  & 1.01 (0.73, 1.29)    & 0.27 (0.16, 0.38)          & 0.81 & 4.95\\
    & jMRUI QUEST  & 1.06 (0.72, 1.39)    & 0.28 (0.15, 0.41)          & 0.77 & 5.52\\
    & LCModel      & 0.24 (0.21, 0.27)    & 0.013 (0.00068, 0.025)     & 0.95 & 4.33\\
    & Gannet       & 0.67 (0.61, 0.73)    & -0.043 (-0.067, -0.020)    & 0.98 & 2.85\\
    & LWFIT        & 0.76 (0.69, 0.84)    & 0.0067 (-0.023, 0.037)     & 0.97 & 4.22\\
    \hline
    E3 & TARQUIN   & 1.84 (1.72, 1.97)    & -0.046 (-0.12, 0.024)      & 0.99 & 5.55\\
    & jMRUI AQSES  & 1.40 (0.87, 1.93)    & 0.21 (-0.084, 0.50)        & 0.71 & 3.59\\
    & jMRUI QUEST  & 1.42 (1.04, 1.79)    & 0.23 (0.025, 0.44)         & 0.84 & 3.31\\
    & LCModel      & 0.56 (0.55, 0.65)    & -0.044 (-0.07, -0.019)     & 0.98 & 3.77\\
    & Gannet       & 0.56 (0.51, 0.69)    & -0.017 (-0.065, 0.031)     & 0.94 & 3.31\\
    & LWFIT        & 0.98 (0.88, 1.07)    & 0.0067 (-0.023, 0.037)     & 0.97 & 2.60\\
    \hline
    E4 & TARQUIN   & 1.77 (-0.54, 4.08)   & 0.13 (-0.69, 0.96)         & 0.37 & 42.1\\
    & jMRUI AQSES  & 1.17 (0.15, 2.18)    & 0.48 (0.12, 0.84)          & 0.57 & 5.85\\
    & jMRUI QUEST  & 1.02 (-0.18, 2.22)   & 0.52 (0.092, 0.95)         & 0.42 & 6.11\\
    & LCModel      & 0.45 (0.22, 0.69)    & 0.036 (-0.048, 0.12)       & 0.79 & 7.83\\
    & Gannet       & 0.51 (0.34, 0.69)    & 0.086 (0.024, 0.15)        & 0.90 & 2.82\\
    & LWFIT        & 0.91(0.55, 1.28)     & 0.051 (-0.081, 0.18)       & 0.86 & 2.29\\
    \hline
  \end{tabular}
\end{table}

The quantification results obtained by the various fitting approaches are summarised in \Cref{Tab:Experimental}. The results are reported as gradient, $a$, and intercept, $b$, with $95\%$ confidence intervals and $R^2$ value of the linear fit. The concentration ratios are also plotted against the known ground-truth ratios in \Cref{fig:E1}--\Cref{fig:E4}.

\Cref{fig:E1} shows the GABA-to-NAA ratio for the pH calibrated solution series without Glx, E1. LWFIT is able to achieve a high fidelity fit for this series, with a GABA gradient estimate $0.98$ with $R^2$ of $0.97$. TARQUIN, on the other hand, overestimates $a$ by almost a factor of three. All other tools underestimate the GABA gradient to varying degrees, from $0.56$ for Gannet to $0.67$ for AQSES. The $R^2$ values of the linear fit are generally good, with TARQUIN, LCModel, Gannet and LWFIT all producing values greater than $0.87$. While the two jMRUI tools produce a lower fit fidelity with an offset, $b$, of around $0.5$, they are at least consistent with one and other.

\Cref{fig:E2} shows results for the intentionally miscalibrated solution series, E2. Unlike the other series, E2 did not exhibit the characteristic pseudo-triplet GABA resonance, and instead presented a single broad peak at \SI{3}{ppm}. While a pH of $3.0$ is beyond what is commonly observed \invivo{}, this series was intended as a means to study the effect of chemical environmental changes upon quantification, a relevant consideration when investigating certain pathologies. Surprisingly, both TARQUIN and jMRUI give significantly more accurate GABA-to-NAA gradients, with $a$ close to $1$ and higher $R^2$ than for the pH-calibrated series. The TARQUIN change appears to be a consequence of its previous tendency to over-estimate GABA, while jMRUI's GABA gradient increase is likely driven by a reduced efficacy in NAA quantification. Peak fitting tools were expected to be more robust to environmental pH changes. With less tightly constrained models, the majority of signal is still captured by the tool, leaving only a systematic offset resulting from edit efficiency correction. We found that LWFIT and Gannet maintain a high $R^2$, with LWFIT showing a reduced GABA-to-NAA gradient, likely a result of integration limits cutting off part of the broad resonance. Gannet, on the other hand, exhibits a modest increase, however there is significant overlap with the E1 confidence interval, so this finding may not be significant. Gannet's Gaussian model of GABA seems a suitable choice for this series, and the robustness of Gannet's estimations support this.

\Cref{fig:E3} shows the results for the pH-calibrated solutions with Glu and Gln.  TARQUIN, LCModel, LWFIT and Gannet largely mirror their respective results for E1. TARQUIN overestimates the gradient, LCModel and Gannet estimate an $a$ of around $0.5$--$0.6$, and LWFIT performs well, with an estimated gradient of $0.98$ and $R^2$ of $0.97$. The quality of the GABA-to-NAA fits is better across the board, with higher $R^2$ values ranging from $0.71$ to $0.99$ for jMRUI and TARQUIN, respectively. The overall improvement in quality of the linear fits is unlikely to be related to the introduction of Glu and Gln, but could be due to a small change in the experimental procedure resulting in fewer and smaller adjustments in transmitter settings and shim gradients between scans as illustrated in \Cref{tab:calibration}.

\Cref{fig:E4} shows the results for the gel phantom series with Glu and Gln. The fit quality for the gel series is generally lower, as expected due to higher FWHM and lower SNR for tissue-mimicking gels. The mean $R^2$ of E4 GABA-to-NAA gradient estimates across all tools is $0.65$ for E4, compared to $0.85$, $0.90$ and $0.91$ for E1-E3, respectively. Despite this, there is significant overlap of confidence intervals of E3 and E4 GABA-to-NAA gradient estimates for all quantification methods, suggesting that this drop in data quality does little to compromise the quantification.

\section{Discussion}

The main aim of this work was to develop an iterative phantom-based method for benchmarking GABA quantification software. The phantom data exhibit a high SNR with narrow metabolite line-widths, well above quality thresholds outlined in consensus work. The four experimental series provide unique challenges to quantification software, allowing elucidation of several confounds which would not be possible using \invivo{} data, alone. E1 represents the simplest case with only GABA, NAA and Cr, E2 introduces environmental changes to pH, E3 adds Glx to investigate its potential to interfere in GABA quantification, and the gels of E4 present broader and lower SNR spectra. LWFIT --- intended to serve as a baseline sanity check --- reproduces the expected gradient to a high fidelity for E1, E3 and E4. LWFIT does, however, underestimate the GABA-to-NAA concentration gradient for E2, with its pH offset. The broad \SI{3}{ppm} GABA resonance present in E2 is likely insufficiently captured by the integration limits used by LWFIT, furthermore, edit efficiency calculations are likely invalidated somewhat by the apparent structural changes in the spectra.

While all tools generally estimated a reasonable linearity in the GABA-to-NAA gradients, there is a systematic offset in these estimates. For TARQUIN, jMRUI and LCModel, these changes are primarily driven by the choice of basis set. This can be seen by comparing the two jMRUI methods --- QUEST and AQSES --- which share a common basis. Their mean estimates of GABA-to-NAA gradients differed by just $4\%$ across all the experimental series. The discrepancy increases to around $106\%$ when comparing TARQUIN and LCModel. The basis issue is further complicated by differences in the default GABA models used by the various tools. LCModel includes the Govindaraju and Kaiser model, while jMRUI uses its own modified version of the Govindaraju and Kaiser models. GABA spectra simulated using the Govindaraju, Kaiser and Near model, shown in \Cref{fig:Sim_Models} exhibit clear differences in the structure of the \SI{3}{ppm} GABA peak. Generally, the Kaiser and Near models appear to approximate the experimental spectra better than the older Govindaraju model, in agreement with observations in the literature. However, despite this, many tools still maintain the older Govidaraju GABA model by default. It is therefore not possible to fundamentally ascertain the performance of these tools without a systematic study with common basis sets and optimised settings. An interesting follow-up study would be to generate a common optimised basis set to be used in several basis set fitting methods. The phantom series can be further enriched by adding degrees of noise, line-broadening, experimentally acquired macromolecular baselines and other co-edited resonances to investigate their influence on quantification, providing many avenues of further study.

With the exception of TARQUIN, the tools present a tendency to underestimate GABA-to-NAA ratios to varying degrees. Simulations suggest realistic refocusing pulses reduce the editing efficiency and can lead to a reduction in reported GABA-to-NAA ratios of up to $20\%$. Including additional losses in experiments due to line broadening as a result of imperfect shimming, and excitation pulses with realistic slice profiles could therefore explain the underestimation of GABA-to-NAA gradients by many of the tools discussed. These losses can be taken into account by appropriate simulation during the creation of basis functions, or as a part of the edit efficiency correction for peak fitting tools.

\begin{table}
  \caption{Mean, $\bar{a}$, and standard deviations, $\sigma(a)$, of GABA-to-NAA gradient estimates for all tools.}
  \label{Tab:MetaExp}
  \centering
  \begin{tabular}{|c|cccccc|}
    \hline
    Tool & TARQUIN & jMRUI AQSES & jMRUI QUEST & LCModel & Gannet & LWFIT\\
    \hline
    $\bar{a}$   & 1.88 & 1.06 & 1.02 & 0.48 & 0.58 & 0.91\\
    $\sigma(a)$ & 0.73 & 0.31 & 0.35 & 0.18 & 0.07 & 0.10\\
    \hline
  \end{tabular}
\end{table}

\Cref{Tab:MetaExp} contains a meta analysis of the reported GABA-to-NAA gradient values, specifically the mean gradient estimate and standard deviation of this value across all experimental series. The mean provides an indication of each tool's tendency to under or over estimate the changes in GABA-to-NAA. Whereas the standard deviation, $\sigma$, provides an indication of the robustness of each tool's estimations to changes in the background conditions. A notable observation is that the peak fitting methods --- Gannet and LWFIT --- have far lower $\sigma$ than all basis set methods. This suggests that the less strictly parametrised approaches might be preferable for edited spectroscopy analysis, where the landscape is greatly simplified compared to conventional MRS. This is in agreement with the observations of Marzola~\etal{}~\cite{Marzola2014}, who also reported the robustness of these methods \emph{in-vitro} but note the poor performance of integration methods \invivo{}. However, it should be noted that when series E2 is excluded, LCModel exhibits a similar results to Gannet, with a mean and standard deviation of $0.5583$ and $0.0939$, respectively.

The fit error is one aspect that was not approached in this study. While basis set methods generally report the Cramer-Rao lower-bound (CRLB) as the preferred measure of fit quality for a particular basis signal, this functionality is not present in Gannet and not valid for LWFIT, limiting comparability. Furthermore, the use of CRLB as a quality filter has been shown to cause bias in mean concentration estimates~\cite{kreis2016CRLB}. Fit residuals could provide some insight into fit accuracy, however these will be modulated by baseline approximation procedures, which differ substantially between tools, making comparisons potentially unfair. While efforts were made to ensure reasonable fit quality in this study, future comparative work will need to establish a means of common error reporting, using the same, sensibly justified metric. This is a fundamental consideration of quantification benchmarking.

\section{Conclusion}

The phantom benchmark dataset presented here provided a novel, interesting way to compare GABA MRS analysis software. The dataset is intended to form part of a larger systematic benchmarking of analysis tools, a necessary future step for the MRS community. While indeed preliminary, the initial comparison of software seems to indicate that a less-strictly parametrised peak fitting methods might be the best option for edited MRS, where the simplified spectral landscape does not necessitate complete modelling of the peak. This is, however, still a necessary step, as modelling of edit efficiency is important in reducing systematic offsets in this approach. The methodological dependence of the quantification results observed also suggests that care must be taken when comparing results across studies, where analysis pipelines may differ and under-reporting of fitting methods muddy the waters for study replication.

\section*{Acknowledgements}

The authors would like to thank the Experimental MRI Center (EMRIC, Cardiff University) for the use of their LCModel license. This project received funding from: Royal Society Leverhulme Trust Senior Fellowship (SMS); ABMU Health Board, College of Science, Welcome trust strategic award - 104943/Z/14/Z (CJ); EPSRC studentship (MC)


\subsection*{Financial disclosure}

None reported.

\subsection*{Conflict of interest}

The authors declare no potential conflict of interests.



\bibliography{BenchmarkingGABA}

\appendix

\section{Supplementary material}

\begin{figure}[!htb]
  \begin{subfigure}{.414\textwidth}
    \includegraphics[width=1\linewidth]{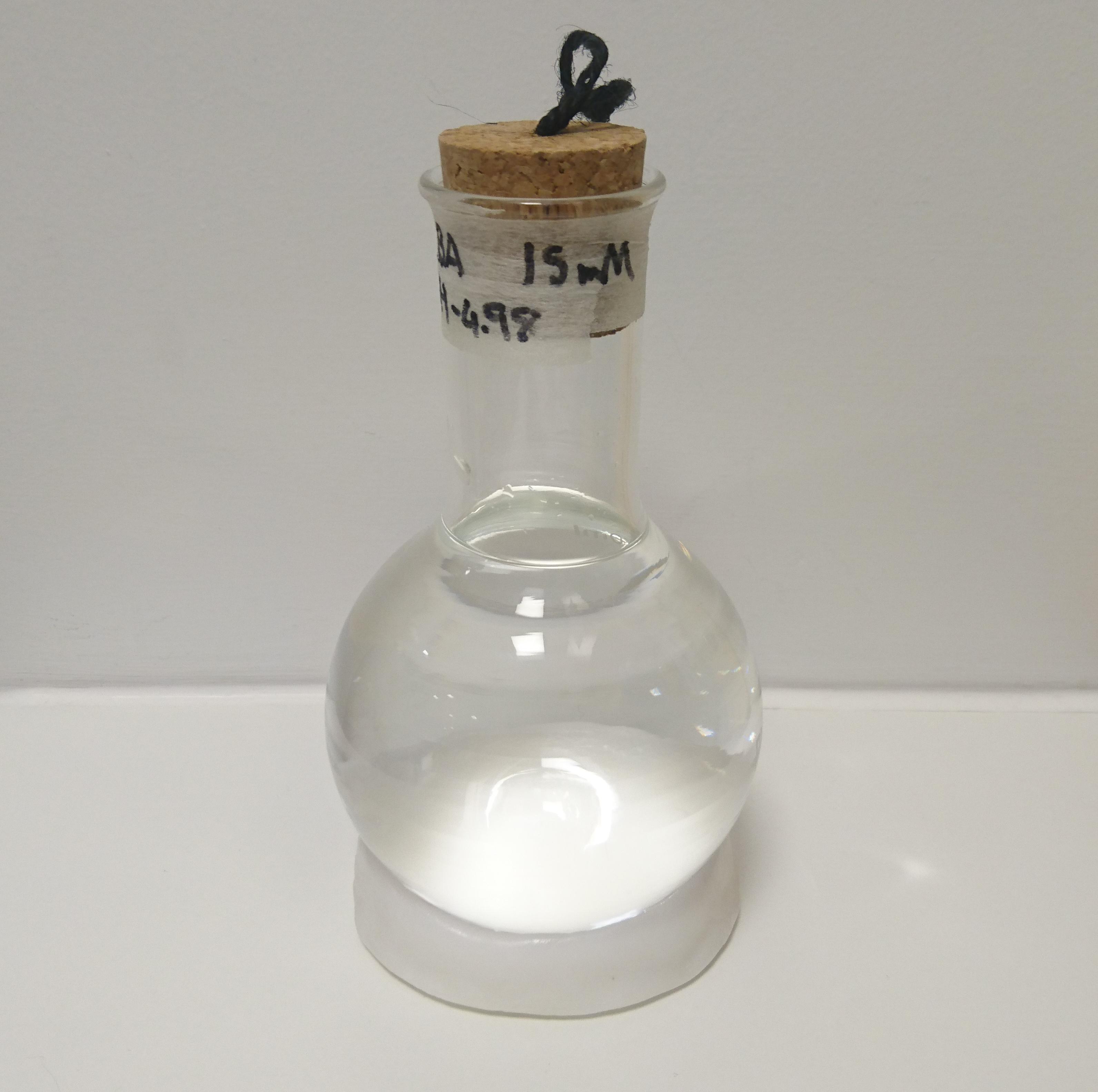}
    \subcaption{Solution phantom.}
  \end{subfigure}\hfill
  \begin{subfigure}{.5675\textwidth}
    \includegraphics[width=1\linewidth]{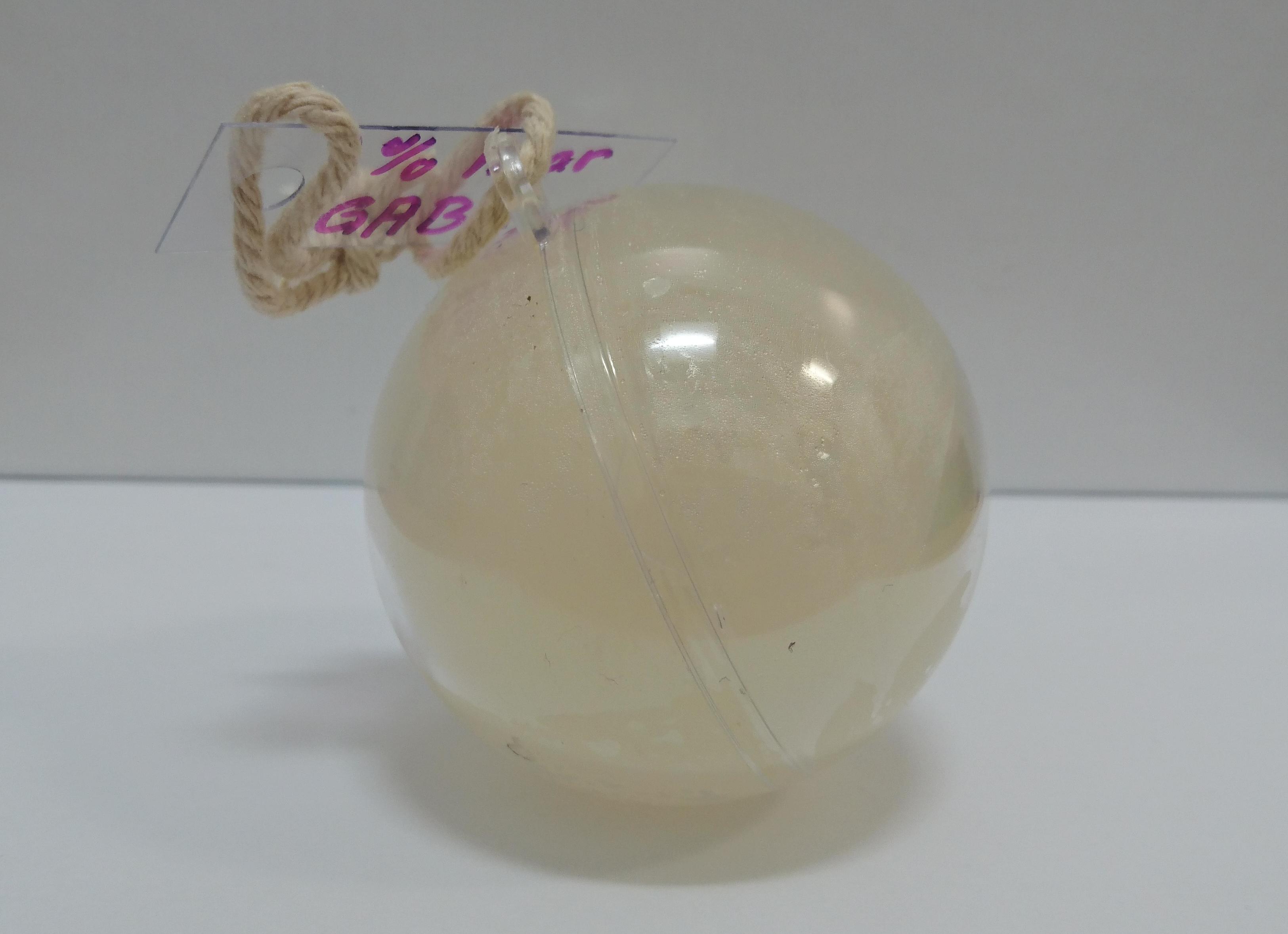}
    \subcaption{Spherical gel phantom.}
  \end{subfigure}
  \caption{Images of phantoms. The gel phantoms are approximately \SI{5.8}{\centi\metre} in diameter and the diameter of the flask is approximately \SI{8.2}{\centi\metre}.}\label{Fig:PhantomPics}
\end{figure}

\begin{figure}[!htb]
  \begin{subfigure}{.32\textwidth}
    \includegraphics[width=1\linewidth]{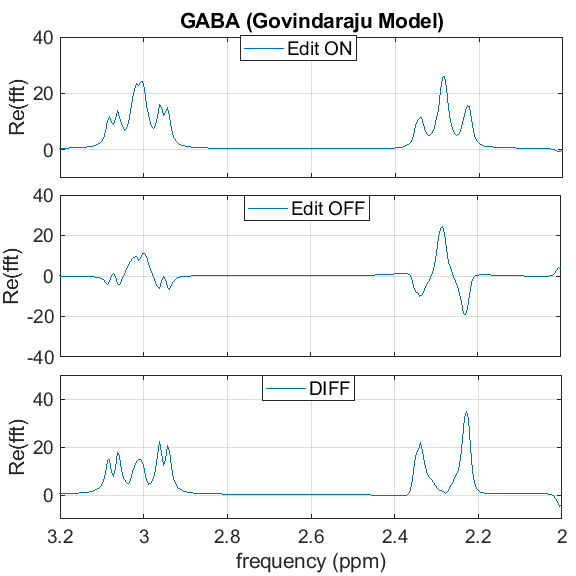}
    \subcaption{Govindaraju \etal{}}
  \end{subfigure}\hfill
  \begin{subfigure}{.32\textwidth}
    \includegraphics[width=1\linewidth]{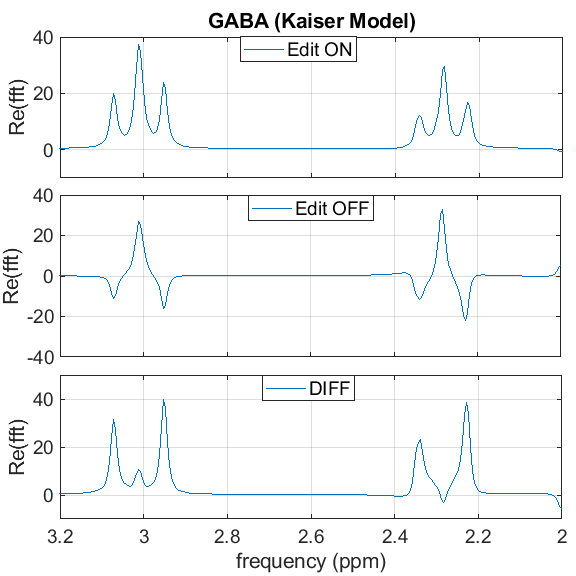}
    \subcaption{Kaiser \etal{}}
  \end{subfigure}\hfill
  \begin{subfigure}{.32\textwidth}
    \includegraphics[width=1\linewidth]{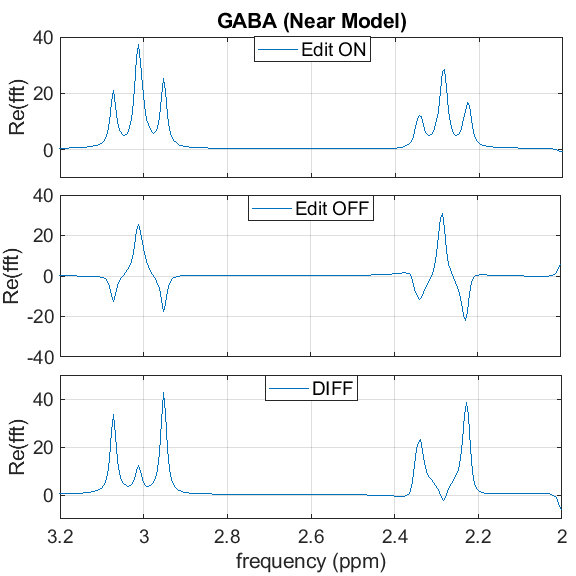}
    \subcaption{Near \etal{}}
  \end{subfigure}\hfill
  \caption{Simulated GABA spectra for the three models considered. Kaiser \etal{} and Near \etal{} models differ only in some couplings, whereas Govindaraju's model produces a significantly different spectrum.}
  \label{fig:Sim_Models}
\end{figure}

\end{document}